\documentclass{article}

\usepackage{arxiv}

\usepackage[utf8]{inputenc} 
\usepackage[T1]{fontenc}    
\usepackage{hyperref}       
\usepackage{url}            
\usepackage{booktabs}       
\usepackage{amsfonts}       
\usepackage{nicefrac}       
\usepackage{microtype}      
\usepackage{lipsum}		
\usepackage{graphicx}
\usepackage{natbib}
\usepackage{doi}
\usepackage{amsmath}

\usepackage{fancyhdr}
\usepackage{textcomp}

\title{A Dynamic Coupling Theory of Expertise Through Thinking Flow and Workflow Evolution}

\author{ 
	\href{https://orcid.org/0009-0004-1760-0149}{\includegraphics[scale=0.06]{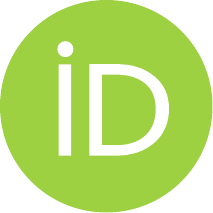}\hspace{1mm}Annie Yuan} \\
	School of Computer Science\\
	The University of Sydney\\
	NSW, 2006, Australia \\
	\texttt{annie.yuan@sydney.edu.au} \\
}

\renewcommand{\shorttitle}{A Dynamic Coupling Theory of Expertise}

\hypersetup{
pdftitle={Tacit Signal Infrastructure: Towards AI Systems that Model Expert Sensing Over Time},
pdfsubject={Artificial Intelligence, Human--AI Interaction, Expert Cognition, Cognitive Infrastructure},
pdfauthor={Annie Yuan},
pdfkeywords={Tacit Signal Infrastructure, Expert Weak Signal Perception, Weak Cognitive Signals, Tacit Signal Computability, Longitudinal Cognitive Operations, Expert Cognition, Weak Signals}
}

\begin{document}
\maketitle
\thispagestyle{plain}

\begin{abstract}
For more than half a century, expertise has predominantly been explained through the lenses of tacit knowledge, deliberate practice, skill acquisition, and expert performance. Whilst these perspectives have contributed significantly to our understanding of expertise, they largely describe its outcomes rather than the underlying cognitive architecture through which expertise continuously emerges and evolves. Consequently, a fundamental question remains insufficiently addressed: what dynamic cognitive mechanism generates expertise across time, context, and professional practice?
This paper proposes Workflow Cognition as a new theoretical framework for understanding expertise. Workflow Cognition is defined as the dynamic cognitive architecture emerging from the recursive coupling between Thinking Flow and Workflow Evolution. Thinking Flow refers to the ongoing processes of perception, interpretation, judgement, decision-making, and reflection, whilst Workflow Evolution refers to the continuous adaptation and transformation of actions, task structures, and operational strategies within situated practice. Through their dynamic coupling, expertise is not accumulated as a static body of knowledge, but continuously generated as an evolving cognitive phenomenon.
Building upon this framework, the paper advances a new ontological definition of expertise. Rather than conceptualising expertise as knowledge, skill, or performance, expertise is defined as an emergent manifestation of Workflow Cognition operating across longitudinal professional experience. Knowledge, skills, decisions, aesthetic preferences, and behavioural patterns are therefore interpreted as observable expressions of expertise rather than expertise itself.
Drawing upon comparative analyses across craft, creative, educational, and leadership domains, the paper introduces a Dynamic Coupling Model of Expertise and establishes the theoretical foundation for subsequent investigations into Longitudinal Tacit Cognition, Longitudinal Aesthetic Cognition, and Expertise Workflow Grammar. The framework contributes a new cognitive ontology of expertise and provides a foundation for future computational representations of human expertise within AI+Expert systems.
\end{abstract}


\section{Introduction}
Expertise has long occupied a central position in research on cognition, learning, professional development, and human performance. Across domains such as craftsmanship, design, education, leadership, medicine, and the creative industries, experts consistently demonstrate capabilities that distinguish them from novices. They recognise subtle patterns, make effective decisions under uncertainty, anticipate emerging situations, and generate solutions that may appear intuitive or inaccessible to less experienced practitioners.

Despite decades of research, a fundamental question remains unresolved: why are experts different from novices? Although expert performance can be observed, measured, and evaluated, the cognitive mechanisms responsible for the emergence and evolution of expertise remain difficult to explain. In many cases, experts themselves struggle to articulate how they arrive at particular judgements or decisions. Their most valuable capabilities often appear to operate beneath conscious verbalisation and therefore resist formal description.

This challenge becomes particularly significant when expertise must be transmitted, taught, or scaled. In domains involving long-term apprenticeship, creative practice, leadership development, or intangible cultural heritage transmission, extensive documentation and instruction often fail to reproduce expert-level cognition in learners. The persistence of this problem suggests that expertise cannot be adequately understood as knowledge possession alone. Rather, expertise may involve a deeper cognitive architecture that continuously organises perception, interpretation, judgement, decision-making, and action throughout real-world practice.

Understanding this architecture is increasingly important not only for expertise research, but also for emerging human--AI systems that seek to represent, transfer, and reconstruct human expertise computationally.

Among existing theories of expertise, Tacit Knowledge Theory has exerted a particularly significant influence. Polanyi's well-known proposition that ``we know more than we can tell'' \cite{polanyi1966logic} shifted scholarly attention towards forms of knowledge that cannot be fully articulated through language. Subsequent research in knowledge management, organisational learning, and professional practice has further demonstrated the importance of tacit knowledge in expert performance \cite{nonaka2007knowledge}.

However, while tacit knowledge explains why expertise is often difficult to articulate, it does not fully explain how expertise is organised and enacted during ongoing practice. Tacit knowledge describes an important characteristic of expertise, namely its partial resistance to explicit representation, but it does not provide a comprehensive account of the cognitive architecture through which expertise continuously emerges, adapts, and evolves.

In particular, Tacit Knowledge Theory leaves several questions unresolved. How are expert judgements organised during real-time practice? How do cognition and workflow influence one another over extended periods of professional development? How do experts adapt their thinking in response to changing situations, constraints, and feedback? Most importantly, what underlying cognitive structure enables expertise to persist and evolve across years, or even decades, of practice?

This paper argues that expertise should not be understood primarily as accumulated knowledge, acquired skill, accumulated experience, or tacit knowledge alone. Instead, expertise is proposed to emerge from an underlying cognitive architecture that continuously organises thinking and action within evolving workflows.

A substantial body of literature has attempted to explain expertise through different theoretical perspectives. Tacit Knowledge Theory emphasises implicit understanding \cite{polanyi1966logic}. Deliberate Practice Theory highlights sustained, goal-directed practice as a primary mechanism underlying expert performance \cite{ericsson1993role}. Skill Acquisition Models describe progressive transitions from novice to expert through increasing contextual judgement \cite{dreyfus1986mind}. Knowledge Management perspectives focus on the creation, transfer, and application of knowledge within organisations \cite{nonaka2007knowledge}. Research on expert performance and naturalistic decision-making examines behavioural outcomes and decision processes among highly experienced practitioners \cite{klein2011expert,chi2014nature}.

While these perspectives provide valuable insights, they share an important limitation. Many explain expertise through possession-based constructs such as knowledge, skill, experience, or tacit understanding. Others focus primarily on observable manifestations, such as performance outcomes and decision quality. As a result, expertise is often treated either as something individuals possess or as something they demonstrate.

What remains insufficiently explained is the dynamic cognitive architecture through which expertise continuously operates. Existing theories rarely account for how perception, interpretation, judgement, decision-making, action, and reflection become recursively organised during ongoing practice. Nor do they adequately explain how cognition and workflow co-evolve over time to generate expertise itself.

Current theories therefore explain important dimensions of expertise, including knowledge accumulation, skill acquisition, expert performance, and practice-based learning. However, they do not adequately explain the cognitive architecture through which expertise continuously emerges and evolves. More specifically, there remains no comprehensive framework capable of explaining how cognition and workflow become dynamically coupled during professional practice, or how this coupling generates the structures commonly recognised as expertise.

This gap is particularly significant for contemporary efforts to computationally represent expertise, design AI-assisted learning systems, and develop scalable approaches to expertise transmission. Without an account of the underlying cognitive architecture of expertise, existing approaches remain limited to modelling observable knowledge, behaviours, or outcomes.

\begin{figure}[htbp]
    \centering
    \includegraphics[width=0.95\linewidth]{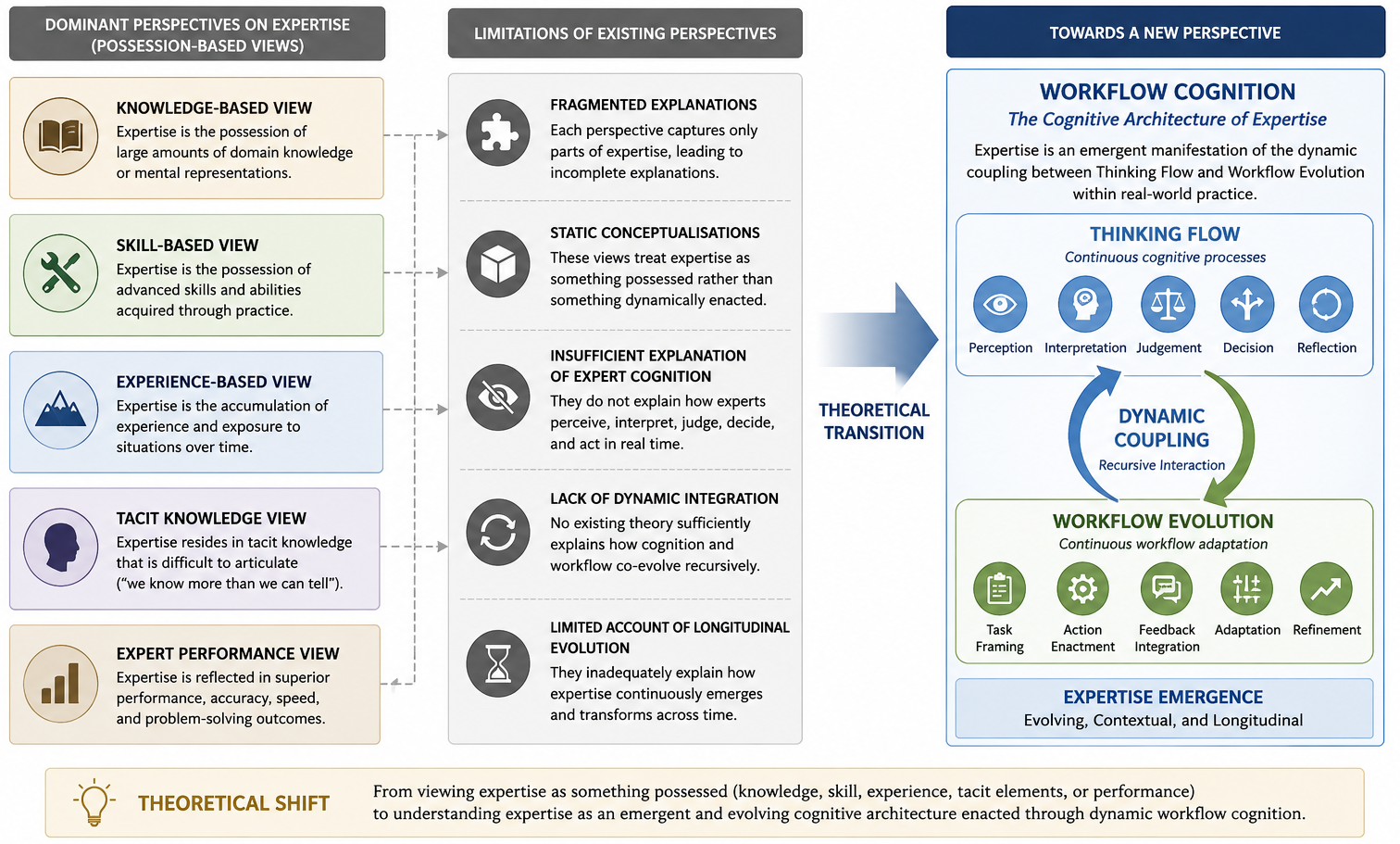}
    \caption{Beyond knowledge, skill, experience, and tacit knowledge: Towards Workflow Cognition as the cognitive architecture of expertise.}
    \label{fig:beyond-knowledge}
\end{figure}

Figure~\ref{fig:beyond-knowledge} summarises the dominant perspectives on expertise and illustrates the theoretical transition proposed in this paper. Rather than viewing expertise as the possession of knowledge, skill, experience, tacit understanding, or performance capabilities, this paper proposes \emph{Workflow Cognition} as the dynamic cognitive architecture underlying expertise.

To address this gap, the paper investigates the following research question:

\begin{quote}
\textbf{RQ:} What cognitive architecture underlies expertise, and how does this architecture organise expert cognition during real-world practice?
\end{quote}

This question shifts the focus of expertise research away from what experts possess towards how expertise operates. Rather than treating expertise as a collection of knowledge, skills, or experiences, the paper seeks to identify the dynamic organisational principles through which expert cognition emerges and evolves.

To address this question, the paper makes three primary contributions:

\begin{enumerate}
    \item \textbf{Workflow Cognition.} The paper introduces \emph{Workflow Cognition} as a theoretical framework for understanding expertise. Workflow Cognition conceptualises expertise as a dynamic cognitive architecture emerging through the continuous interaction of cognition and workflow.

    \item \textbf{Dynamic Coupling Theory.} The paper proposes \emph{Dynamic Coupling Theory}, which explains expertise as an emergent phenomenon arising from the recursive interaction between Thinking Flow and Workflow Evolution. This perspective positions cognition and practice as mutually transformative processes operating across time.

    \item \textbf{A new ontology of expertise.} Building upon Workflow Cognition, the paper advances a new ontological definition of expertise. Rather than viewing expertise as knowledge, skill, experience, or performance, expertise is defined as an emergent manifestation of Workflow Cognition. This ontological perspective establishes the theoretical foundation for subsequent investigations into Longitudinal Tacit Cognition, Longitudinal Aesthetic Cognition, and Expertise Workflow Grammar.
\end{enumerate}

\section{Theoretical Background}

The study of expertise has occupied a central position within cognitive science, education, psychology, organisational learning, and human performance research for several decades. Although different traditions have proposed distinct explanations for expert performance, most seek to understand how individuals acquire, develop, and apply capabilities that distinguish experts from novices. Foundational work in expertise research has shown that experts differ from novices not merely in the amount of knowledge they possess, but also in how they organise, represent, and apply knowledge during complex activity \cite{chi2006two,chi2014nature,ericsson2018cambridge}.

One of the most influential contributions originates from Polanyi's theory of tacit knowledge \cite{polanyi1966logic,polanyi2009tacit}. Polanyi argued that human knowing extends beyond explicit articulation, proposing that individuals frequently know more than they can express verbally. This insight fundamentally reshaped discussions of expertise by highlighting the importance of implicit understanding, intuition, and experiential judgement. Tacit knowledge has since become central to theories of professional practice, organisational learning, and knowledge creation, particularly in accounts of how knowledge is generated, shared, and transformed within practice-based settings \cite{nonaka2007knowledge,argyris1996organizational}.

Subsequent research shifted attention towards expertise acquisition. Ericsson, Krampe, and Tesch-Römer proposed Deliberate Practice Theory, arguing that expert performance emerges primarily through sustained, goal-directed practice conducted over extended periods \cite{ericsson1993role}. Their work challenged assumptions regarding innate talent and emphasised the role of structured learning, feedback, and repeated refinement. Dreyfus and Dreyfus further contributed to expertise research through their Skill Acquisition Model, describing progression from novice to expert as a gradual transition from rule-based behaviour towards context-sensitive judgement \cite{dreyfus1980five,dreyfus1986mind}. Their framework emphasised the increasing importance of intuition, situated understanding, and embodied responsiveness as expertise develops.

Research on expert performance and decision-making has further demonstrated that expertise often operates under conditions of uncertainty, time pressure, and incomplete information. Work on naturalistic decision-making highlights how experts rely on pattern recognition, situational assessment, and experience-based judgement in real-world environments \cite{klein2011expert,klein2017sources}. Similarly, theories of bounded rationality suggest that human decision-making is shaped by constraints, environments, and available information rather than by abstract optimisation alone \cite{simon1972theories,simon2019sciences}. These perspectives are important because they position expertise as a form of adaptive judgement enacted within practical constraints.

Collectively, these contributions have significantly advanced understanding of expertise. However, they also reveal an important pattern. Expertise is frequently explained through knowledge, skill, experience, intuition, or performance outcomes, while the underlying cognitive architecture responsible for organising these phenomena remains comparatively under-theorised.

Among existing theories, Tacit Knowledge Theory remains particularly influential because it directly addresses one of the defining characteristics of expertise: the difficulty of articulation. Polanyi's proposition that ``we know more than we can tell'' has become a foundational assumption across expertise studies, organisational learning, and professional education \cite{polanyi1966logic,polanyi2009tacit}. Tacit knowledge successfully explains several important aspects of expertise. It accounts for why experienced practitioners often struggle to verbalise their decision-making processes, why apprenticeship remains effective in domains such as craftsmanship, medicine, leadership, and creative practice, and why formal documentation frequently fails to capture the full richness of expert understanding.

However, tacit knowledge theory primarily describes the representational characteristics of expertise rather than its organisational mechanisms. It explains why expertise may be difficult to articulate, but provides a more limited account of how expertise operates during ongoing practice. Several questions therefore remain unresolved. How do experts continuously organise perception, interpretation, judgement, and action during real-time workflow execution? How do cognitive processes adapt in response to changing situations and feedback? How does expertise evolve across years of professional engagement rather than simply accumulate?

These questions suggest that tacit knowledge may be a manifestation of a deeper cognitive architecture rather than a complete explanation of expertise itself. This distinction is particularly important for work that seeks to elicit, model, or computationally represent expert knowledge, where the challenge is not only to extract what experts know, but also to understand how expert judgement is organised and enacted in practice \cite{hoffman1987problem,freire2023tacit,cha2023unlocking}.

Additional insights into expertise emerge from situated and practice-based theories of cognition. These perspectives reject the assumption that cognition can be fully understood as an internal process detached from context, activity, and social practice. Lave and Wenger proposed Situated Learning Theory, arguing that learning occurs through participation in communities of practice rather than through abstract knowledge acquisition alone \cite{lave1991situated,wenger1999communities}. Expertise develops through engagement with social, cultural, and practical environments. Similarly, Brown, Collins, and Duguid emphasised that cognition is inseparable from the situations in which it is enacted \cite{brown1989situated}. Their work demonstrated that knowledge acquires meaning through practice and cannot be fully understood outside the contexts in which it is used.

Theories of apprenticeship and professional learning extend this view by emphasising the visibility of expert thinking during practice. Cognitive apprenticeship proposes that learning involves making expert processes visible so that learners can observe, appropriate, and gradually internalise expert strategies \cite{collins1991cognitive,collins2018cognitive}. Rogoff similarly conceptualised cognitive development as apprenticeship in thinking, emphasising guided participation within social and practical contexts \cite{rogoff1990apprenticeship}. These accounts are important because they show that expertise is not merely transmitted as information, but developed through participation in meaningful activity.

Schön further extended these perspectives through his account of the reflective practitioner \cite{schon1987educating,schon2017reflective}. He argued that professional expertise emerges through cycles of action and reflection, enabling practitioners to adapt their understanding in response to complex, uncertain, and changing situations. This emphasis on reflection-in-action is particularly relevant to Workflow Cognition because it foregrounds the continuous interaction between thinking and doing. Related work on practice theory similarly positions knowledge as something enacted through situated forms of activity rather than simply possessed by individuals \cite{schatzki2001practice}.

Distributed and extended cognition perspectives provide a further theoretical foundation for understanding expertise as situated within systems of people, tools, environments, and representations. Hutchins demonstrated that cognition can be distributed across individuals and artefacts within complex work systems \cite{hutchins1995cognition}, while Pea and Salomon highlighted the educational significance of distributed intelligence and distributed cognition \cite{pea1993practices,salomon1997distributed}. Hollan, Hutchins, and Kirsh later argued that distributed cognition offers a foundation for human--computer interaction research because cognitive activity is often organised across internal processes, external representations, and technological infrastructures \cite{hollan2000distributed}. These perspectives are relevant because they suggest that expert cognition cannot be fully understood by examining the individual mind alone; it must also be understood through interaction with workflows, tools, materials, environments, and social systems.

Together, situated, practice-based, reflective, and distributed cognition perspectives provide an important shift away from static conceptions of expertise. They highlight the dynamic relationship between cognition, context, action, tools, feedback, and learning. Nevertheless, while these theories successfully explain the contextual and enacted nature of expertise, they do not fully specify the cognitive architecture through which expertise becomes organised, sustained, and transformed across time.

The preceding review reveals a common limitation across existing expertise theories. Tacit knowledge explains why expertise is difficult to articulate. Deliberate practice explains how expertise may develop through repeated training. Skill acquisition models explain transitions between levels of competence. Naturalistic decision-making explains expert judgement under uncertainty. Situated learning explains the contextual nature of expertise. Reflective practice explains adaptive professional learning. Distributed cognition explains how cognitive activity is organised across people, artefacts, and environments.

However, none of these perspectives explicitly identifies the cognitive architecture through which expertise continuously operates. A recurring assumption across these theories is that cognition and practice interact. Yet the mechanisms governing this interaction remain insufficiently specified. Existing theories rarely explain how perception, interpretation, judgement, decision-making, action, feedback, and adaptation become recursively organised during ongoing workflow execution.

This paper argues that expertise should be understood as an emergent phenomenon arising from the continuous interaction between cognition and workflow. Rather than treating cognition and practice as separate domains, they should be viewed as dynamically coupled processes that continuously shape one another. To address this theoretical gap, the next section introduces \emph{Workflow Cognition} as a dynamic cognitive architecture through which expertise emerges, operates, and evolves during real-world practice.

\section{Workflow Cognition}

The preceding review suggests that existing theories of expertise explain important dimensions of expert practice, yet none adequately specifies the cognitive architecture through which expertise continuously operates. This paper therefore proposes \emph{Workflow Cognition} as a theoretical framework for understanding expertise.

Workflow Cognition shifts the focus of expertise research away from questions of knowledge possession and towards questions of cognitive organisation. Rather than asking what experts know, Workflow Cognition seeks to explain how expert cognition is continuously organised, enacted, and refined through ongoing interaction with professional workflows.

\begin{figure}[htbp]
    \centering
    \includegraphics[width=0.95\linewidth]{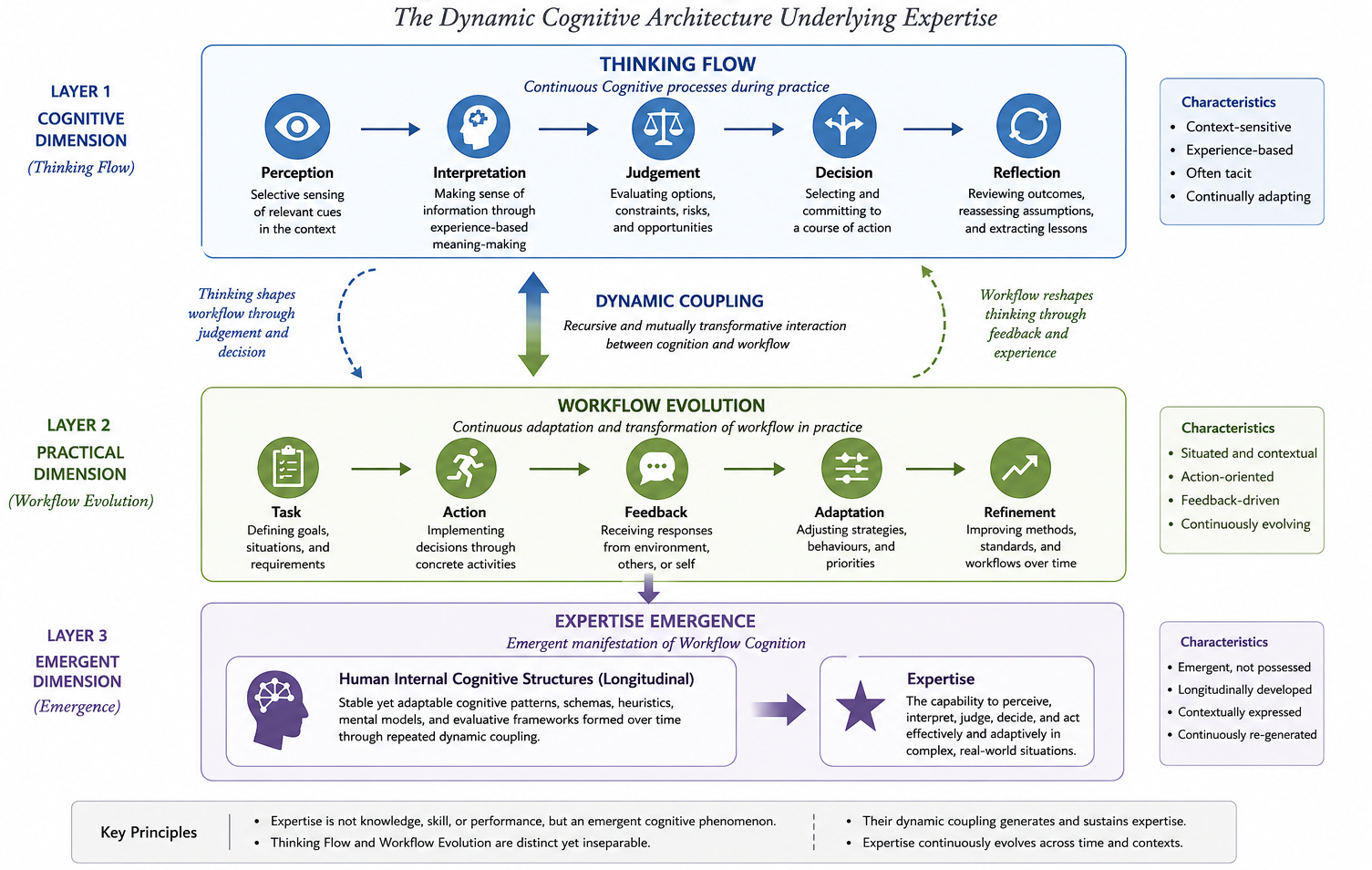}
    \caption{Workflow Cognition architecture: The dynamic cognitive architecture underlying expertise.}
    \label{fig:workflow-cognition}
\end{figure}

The proposed framework conceptualises expertise as an emergent phenomenon arising from the dynamic coupling between cognitive processes and workflow processes. Figure~\ref{fig:workflow-cognition} presents the overall architecture of Workflow Cognition.

\paragraph{Definition 1: Workflow Cognition.}
\textbf{Workflow Cognition} is the dynamic cognitive architecture through which expertise emerges from the recursive coupling of \emph{Thinking Flow} and \emph{Workflow Evolution}.

Under this definition, expertise is not treated as a static possession of knowledge, skill, or experience. Instead, expertise is understood as an ongoing process through which cognition and workflow continuously shape one another.

Workflow Cognition consists of two interdependent components. The first is \textbf{Thinking Flow}, which represents the continuous cognitive processes involved in perception, interpretation, judgement, decision-making, and reflection. The second is \textbf{Workflow Evolution}, which represents the continuous adaptation and transformation of tasks, actions, strategies, and operational structures during practice. Neither component alone is sufficient to explain expertise. Expertise emerges through their continuous interaction.

\paragraph{Definition 2: Thinking Flow.}
\textbf{Thinking Flow} refers to the continuous cognitive process through which individuals perceive, interpret, judge, decide, and reflect during task execution.

Traditional theories of expertise frequently treat cognition as an internal mental activity occurring prior to action. Workflow Cognition adopts a different perspective. Thinking is viewed as a continuous process operating throughout professional practice rather than preceding it.

Five major elements constitute Thinking Flow:

\begin{description}
    \item[\textbf{Perception.}] Perception involves the selective recognition of relevant signals within an environment. Experts often perceive patterns, anomalies, and opportunities that remain invisible to novices. Such perception extends beyond sensory observation and includes the recognition of contextual significance.

    \item[\textbf{Interpretation.}] Interpretation refers to the process through which perceived information acquires meaning. Experts organise observations within broader cognitive structures developed through experience, enabling rapid situational understanding.

    \item[\textbf{Judgement.}] Judgement involves the evaluation of alternatives, constraints, risks, and opportunities. Expert judgement frequently incorporates tacit considerations that are difficult to fully articulate yet play a decisive role in action selection.

    \item[\textbf{Decision.}] Decision-making represents the transition from cognitive evaluation to operational commitment. Decisions determine subsequent actions and shape the trajectory of workflow execution.

    \item[\textbf{Reflection.}] Reflection enables experts to evaluate outcomes, reassess assumptions, and refine future actions. Through reflection, cognition becomes adaptive and self-correcting across time.
\end{description}

Together, these processes form a continuous cognitive flow that operates throughout professional practice.

\paragraph{Definition 3: Workflow Evolution.}
\textbf{Workflow Evolution} refers to the continuous adaptation and transformation of tasks, actions, operational strategies, and procedural structures throughout practice.

While Thinking Flow represents the cognitive dimension of expertise, Workflow Evolution represents its practical dimension. Workflows are not static sequences of actions. They continuously evolve in response to context, feedback, constraints, and learning.

Five major elements constitute Workflow Evolution:

\begin{description}
    \item[\textbf{Task.}] Tasks provide the operational context within which expertise is enacted. Different tasks activate different cognitive structures and decision requirements.

    \item[\textbf{Action.}] Actions represent the observable implementation of decisions. Through action, cognition becomes materially expressed within practice.

    \item[\textbf{Feedback.}] Feedback emerges from the consequences of action. It may originate from environmental responses, social interactions, performance outcomes, or self-evaluation.

    \item[\textbf{Adaptation.}] Adaptation refers to adjustments made in response to feedback. Experts continuously modify strategies, priorities, and behaviours in order to maintain effectiveness under changing conditions.

    \item[\textbf{Refinement.}] Refinement represents the cumulative improvement of workflow structures over time. Through repeated cycles of adaptation, workflows become increasingly sophisticated and context-sensitive.
\end{description}

Workflow Evolution therefore describes the practical trajectory through which expertise develops and operates.

\paragraph{Definition 4: Dynamic Coupling.}
\textbf{Dynamic Coupling} refers to the recursive and mutually transformative interaction between Thinking Flow and Workflow Evolution.

The central proposition of Workflow Cognition is that expertise cannot be explained through cognition alone, nor through workflow alone. Rather, expertise emerges through their continuous interaction. Thinking Flow shapes Workflow Evolution by influencing perception, interpretation, judgement, and decision-making. Workflow Evolution simultaneously reshapes Thinking Flow through action, feedback, adaptation, and refinement.

This relationship is recursive rather than linear. Every decision alters subsequent workflows, and every workflow experience influences future cognition. Over time, repeated cycles of interaction generate increasingly complex cognitive structures and behavioural capabilities.

The dynamic coupling between Thinking Flow and Workflow Evolution can be summarised in the following propositions:

\begin{description}
    \item[\textbf{Proposition 1.}] Thinking Flow influences Workflow Evolution.

    \item[\textbf{Proposition 2.}] Workflow Evolution reshapes Thinking Flow.

    \item[\textbf{Proposition 3.}] Expertise emerges from the recursive coupling of Thinking Flow and Workflow Evolution.
\end{description}

Dynamic Coupling therefore serves as the generative mechanism underlying Workflow Cognition.

\paragraph{Towards a new ontology of expertise.}
The introduction of Workflow Cognition necessitates a reconsideration of what expertise fundamentally is. Most existing theories conceptualise expertise through one of three dominant perspectives.

\begin{description}
    \item[\textbf{Knowledge-based views.}] Expertise is often understood as the possession of extensive domain knowledge and sophisticated mental representations.

    \item[\textbf{Skill-based views.}] Expertise is frequently described as the possession of highly developed procedural skills acquired through practice.

    \item[\textbf{Performance-based views.}] Expertise is often evaluated through observable performance outcomes, including speed, accuracy, effectiveness, and decision quality.
\end{description}

While each perspective captures important aspects of expertise, none adequately explains the generative mechanism responsible for expertise itself. Knowledge may support expertise, but knowledge alone does not explain how expertise continuously adapts to novel situations. Skills may facilitate performance, but skills alone do not explain how experts generate new solutions under uncertainty. Performance may reveal expertise, but performance itself remains an outcome rather than an underlying mechanism.

Workflow Cognition therefore proposes a different ontological position.

\paragraph{Definition 5: Ontology of Expertise.}
\textbf{Expertise} is an emergent manifestation of Workflow Cognition, continuously generated through the dynamic coupling of Thinking Flow and Workflow Evolution across time.

Under this definition, expertise is not a possession but an ongoing cognitive phenomenon. It emerges through the recursive organisation of cognition and practice and continuously evolves throughout professional engagement.

This perspective leads to four further propositions:

\begin{description}
    \item[\textbf{Proposition 4.}] Knowledge is not expertise. Knowledge represents one possible expression of expertise but does not constitute expertise itself.

    \item[\textbf{Proposition 5.}] Skills are not expertise. Skills are operational manifestations of expertise rather than its underlying source.

    \item[\textbf{Proposition 6.}] Expert performance is not expertise. Performance provides observable evidence of expertise but should not be confused with the cognitive architecture generating it.

    \item[\textbf{Proposition 7.}] Knowledge, skills, decisions, judgements, aesthetic preferences, and behavioural patterns are observable expressions of expertise rather than expertise itself.
\end{description}

Consequently, expertise should be understood not as a collection of possessions, but as a continuously evolving cognitive system generated through Workflow Cognition.

\section{Dynamic Coupling Model of Expertise}

\begin{figure}[htbp]
    \centering
    \includegraphics[width=0.95\linewidth]{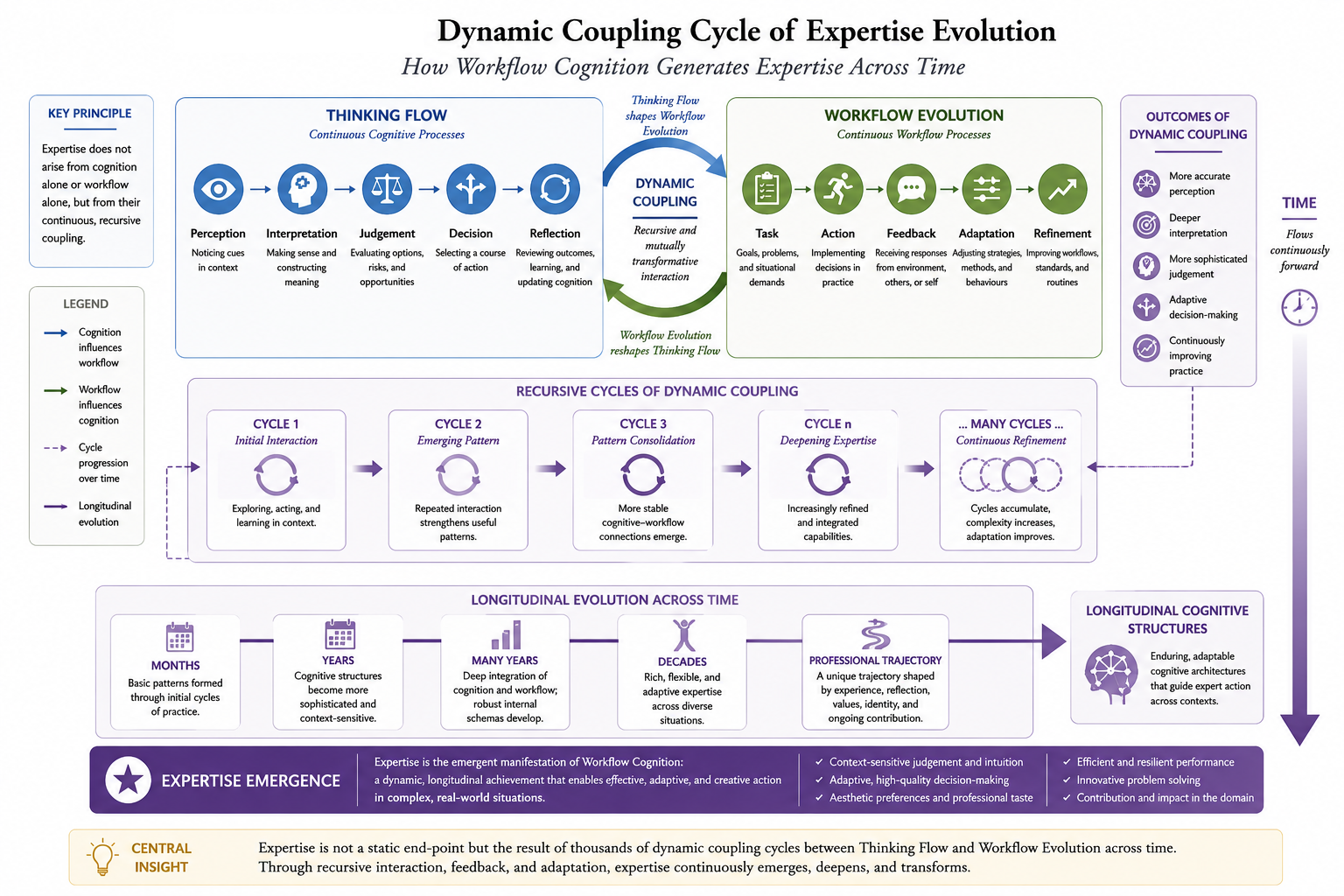}
    \caption{Dynamic Coupling Cycle of Expertise Evolution - How Workflow Cognition generates Expertise Across Time}
    \label{fig:dynamic-coupling-cycle}
\end{figure}

The preceding section introduced Workflow Cognition as the dynamic cognitive architecture underlying expertise. However, the existence of an architecture alone does not explain how expertise continuously emerges, develops, and transforms throughout professional practice.
To address this issue, this section proposes the Dynamic Coupling Model of Expertise. The model explains expertise as a longitudinal phenomenon generated through recursive interactions between Thinking Flow and Workflow Evolution. Rather than viewing expertise as a static achievement or accumulated asset, the model conceptualises expertise as an ongoing process of cognitive and workflow co-evolution.
Figure \ref{fig:dynamic-coupling-cycle} illustrates the Dynamic Coupling Cycle through which expertise continuously develops across time.

\subsection{Recursive Expertise Formation}
Traditional theories frequently describe expertise as the result of accumulated knowledge, prolonged experience, or repeated practice. Whilst these factors undoubtedly contribute to professional development, they do not adequately explain the mechanism through which expertise continuously regenerates itself.
Workflow Cognition proposes that expertise develops through recursive cycles of interaction between cognition and workflow. Each cycle begins with Thinking Flow, during which individuals perceive situations, interpret contextual signals, evaluate alternatives, make decisions, and reflect upon potential actions. These cognitive processes subsequently shape Workflow Evolution through the execution of tasks and actions.
Workflow outcomes generate feedback that returns to the cognitive system. New information, unexpected consequences, and environmental responses influence future perception, interpretation, judgement, and decision-making. Through repeated cycles of interaction, expertise gradually becomes more refined, adaptive, and context-sensitive.
Expertise formation is therefore not linear but recursive. Every cycle modifies both cognition and workflow, creating conditions for subsequent development.

\subsection{Internal Feedback Loops}
The first mechanism supporting Dynamic Coupling consists of internal feedback loops operating within cognition itself.
During professional practice, experts continuously evaluate the consequences of their own decisions and actions. Reflection enables practitioners to compare expected outcomes with actual outcomes, identify discrepancies, and revise future approaches (Schön, 1983).
These internal feedback processes contribute to the development of increasingly sophisticated cognitive structures. Perceptual sensitivity improves. Interpretative frameworks become more nuanced. Judgement becomes more context-aware. Decision-making becomes more adaptive under uncertainty.
Importantly, internal feedback loops do not merely reinforce existing knowledge. They actively reorganise cognitive structures through continual self-correction and refinement.
Consequently, expertise evolves not only through external experience but also through the ongoing reconstruction of internal cognition.

\subsection{External Feedback Loops}
Whilst internal feedback operates within cognition, external feedback originates from interactions between practitioners and their environments.
External feedback may emerge from multiple sources, including:
task outcomes;
environmental responses;
social interactions;
client reactions;
organisational constraints;
market conditions;
cultural contexts.
Each source introduces information that may challenge existing assumptions and require workflow adaptation.
For example, a master craftsperson may encounter unexpected material behaviour. A film director may receive unanticipated responses from audiences. An educator may observe learning outcomes that differ from expectations. In each case, workflow adaptation generates new experiences that subsequently reshape cognition.
External feedback therefore functions as a critical mechanism through which Workflow Evolution influences Thinking Flow. Expertise develops through continuous engagement with environments rather than through isolated cognitive processes.
This perspective aligns with situated theories of cognition whilst extending them through a more explicit account of recursive cognitive-workflow interaction \cite{lave1991situated, brown1989situated} .

\subsection{Longitudinal Evolution}
Perhaps the most significant implication of Dynamic Coupling Theory is that expertise should be understood as a longitudinal phenomenon.
Most expertise theories focus on states of expertise. Dynamic Coupling Theory instead focuses on trajectories of expertise.
Expertise emerges through cumulative cycles of interaction operating across months, years, and often decades of professional practice. Each cycle contributes incremental modifications to both cognition and workflow. Over time, these modifications become organised into increasingly stable yet adaptive cognitive structures.
Longitudinal evolution therefore explains why experienced experts often display forms of judgement that appear intuitive to external observers. Such judgements are not products of isolated decisions but manifestations of thousands of previous cycles of dynamic coupling accumulated throughout professional development.
This perspective also provides a bridge towards the next stage of the AI+Expert theoretical framework. If expertise emerges through longitudinal cycles of cognitive-workflow interaction, then the resulting structures cannot be adequately described solely as tacit knowledge. Rather, they constitute evolving internal cognitive structures formed through long-term professional engagement.
The next paper therefore extends this argument by investigating the emergence of Longitudinal Tacit Cognition, examining how persistent cycles of Workflow Cognition gradually produce the internal cognitive structures underlying expertise.

\section{Comparative Expert Cases}
\begin{figure}[htbp]
    \centering
    \includegraphics[width=0.95\linewidth]{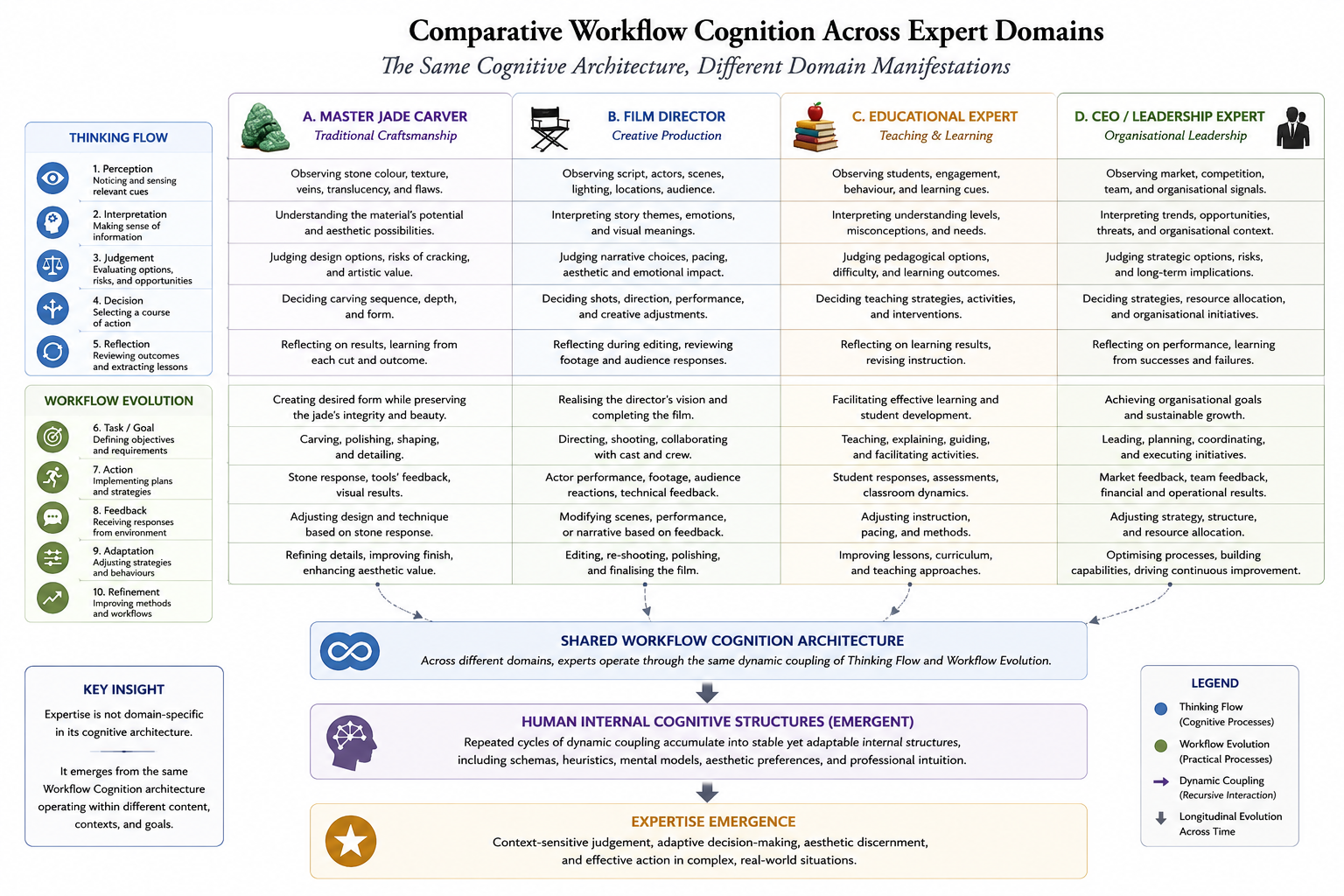}
    \caption{Comparative Workflow Cognition Across Expert Domains - an illustrative example of how the same cognitive architecture applies to different domain manifestations}
    \label{fig:comparative-workflow}
\end{figure}

The preceding sections proposed Workflow Cognition as a dynamic cognitive architecture underlying expertise and explained how expertise emerges through recursive cycles of dynamic coupling. A remaining question, however, concerns the generalisability of the framework.
If Workflow Cognition represents a fundamental architecture of expertise, it should be observable across domains characterised by different tasks, environments, and forms of professional practice. To explore this possibility, this section examines four illustrative expert domains: craftsmanship, creative production, education, and leadership.
The objective is not to compare professional performance itself, but to identify recurring cognitive structures and workflow patterns operating across otherwise distinct forms of expertise.
Figure \ref{fig:comparative-workflow} presents a comparative overview of Workflow Cognition across the selected domains.

\subsection{Master Jade Carver}

Traditional jade carving represents one of the clearest examples of long-term craftsmanship expertise. Master jade carvers routinely make decisions that cannot be fully reduced to explicit rules or technical procedures.

During practice, the expert continuously evaluates the material's colour distribution, internal structure, translucency, texture, and potential imperfections. Perception and interpretation occur simultaneously, and judgements concerning design possibilities often emerge before physical carving begins.

Workflow Evolution occurs through ongoing interaction with the material. Each carving action generates new information regarding the stone's internal characteristics. Feedback from the material continuously reshapes subsequent decisions, modifications, and refinements.

From a Workflow Cognition perspective, expertise does not reside solely in knowledge of jade or in carving skill. Rather, expertise emerges through dynamic coupling between cognitive judgement and workflow adaptation during repeated engagement with the material.

\subsection{Film Director}

Film directing provides an example of expertise operating within highly complex and uncertain environments. Directors must simultaneously interpret scripts, manage actors, coordinate production teams, evaluate visual composition, and respond to constantly changing production constraints. Decision-making therefore occurs under conditions of incomplete information and continuous uncertainty.

Thinking Flow is expressed through ongoing interpretation, aesthetic judgement, creative decision-making, and reflective evaluation. Workflow Evolution occurs through rehearsal, filming, editing, and iterative refinement of the production process.

Importantly, expert directors frequently adjust their creative vision in response to feedback emerging from actors, locations, technical limitations, and audience expectations. Expertise therefore develops not through rigid adherence to predefined plans, but through adaptive interaction between cognition and workflow.

The resulting expertise reflects a continuously evolving cognitive architecture rather than a fixed body of knowledge.

\subsection{Educational Expert}

Educational expertise provides a particularly important case because its primary objective involves facilitating learning in others. Experienced educators continuously monitor learner engagement, interpret student responses, evaluate understanding, and adjust instructional strategies. Effective teaching therefore requires ongoing cycles of perception, interpretation, judgement, decision-making, and reflection.

Workflow Evolution occurs through lesson planning, classroom interaction, assessment, feedback, and instructional adaptation. Every teaching intervention generates information regarding learner progress, which subsequently influences future pedagogical decisions.

Educational expertise is therefore not reducible to subject knowledge alone. Two educators may possess similar disciplinary knowledge yet differ substantially in their ability to facilitate learning.

From a Workflow Cognition perspective, educational expertise emerges through the dynamic coupling between cognitive interpretation of learning situations and the continuous evolution of teaching workflows.

\subsection{CEO and Leadership Expert}

Leadership expertise operates within environments characterised by complexity, ambiguity, and long time horizons. Senior leaders continuously interpret organisational conditions, evaluate opportunities and risks, make strategic decisions, and respond to changing internal and external environments. Their decisions influence not only immediate outcomes but also the future evolution of organisational systems.

Thinking Flow involves strategic perception, contextual interpretation, evaluative judgement, decision-making, and reflective learning. Workflow Evolution occurs through organisational initiatives, resource allocation, stakeholder engagement, implementation processes, and strategic adaptation.

Feedback originates from multiple sources, including employees, customers, competitors, markets, and broader social environments. Effective leaders continuously adapt both their cognitive models and operational approaches in response to this feedback.

Leadership expertise therefore emerges through sustained interaction between cognition and organisational workflows rather than through managerial knowledge alone.

\subsection{Cross-Case Analysis}

Despite substantial differences in professional context, the four cases reveal striking structural similarities.

\paragraph{Shared structures.}
Across all domains, expertise involves:

\begin{itemize}
    \item perception of relevant signals;
    \item interpretation of contextual information;
    \item judgement under uncertainty;
    \item decision-making;
    \item reflection and learning;
    \item workflow adaptation;
    \item feedback integration; and
    \item continuous refinement.
\end{itemize}

These recurring elements correspond directly to the components of Workflow Cognition introduced in Section~\ref{fig:workflow-cognition}. In each case, expertise emerges through dynamic coupling between Thinking Flow and Workflow Evolution rather than through knowledge possession alone.

\paragraph{Different manifestations.}
While the underlying structures appear similar, their manifestations differ substantially across domains. For the jade carver, workflow centres on material interaction. For the film director, workflow centres on creative production. For the educator, workflow centres on learning facilitation. For the leader, workflow centres on organisational adaptation.

Consequently, expertise appears in different forms, produces different outcomes, and operates within different environments. Nevertheless, the underlying architecture remains structurally comparable. This distinction suggests that expertise may possess domain-specific manifestations while simultaneously sharing domain-independent cognitive foundations.

The comparative analysis therefore provides preliminary support for Workflow Cognition as a generalisable framework for understanding expertise across professional contexts. More importantly, it suggests that repeated cycles of Workflow Cognition gradually generate relatively stable internal cognitive structures that persist across time.

The next section discusses the theoretical implications of this finding and its significance for future investigations into longitudinal expertise formation.

\section{Analytic Propositions}

The preceding theoretical analysis and comparative expert cases provide conceptual support for Workflow Cognition as a generalisable framework for understanding expertise. Across domains characterised by different tasks, environments, objectives, and forms of professional practice, several recurring patterns can be identified. These patterns suggest that expertise may be governed by common cognitive principles, despite substantial differences in domain-specific manifestations.

Four analytic propositions are derived from the preceding discussion.

\paragraph{Proposition 1: Workflow Cognition is observable across domains.}
The comparative cases suggest that Workflow Cognition is not restricted to a particular profession or domain. Similar cognitive processes can be identified across craftsmanship, creative production, education, and organisational leadership, despite significant differences in context and operational activity.

In each case, experts continuously engage in perception, interpretation, judgement, decision-making, and reflection while simultaneously adapting workflows in response to feedback and changing circumstances. Although the content of expertise differs substantially between domains, the underlying cognitive architecture displays notable structural similarities.

This proposition suggests that expertise may possess domain-specific manifestations while simultaneously sharing domain-independent cognitive foundations. Workflow Cognition therefore provides a potential framework for understanding expertise beyond disciplinary boundaries. Rather than viewing expertise as a collection of isolated domain-specific competencies, the analysis supports the possibility of a common architecture underlying diverse forms of expert practice.

\paragraph{Proposition 2: Expertise emerges through recursive cognitive--workflow coupling.}
The second proposition concerns the mechanism through which expertise develops and operates. Across the examined domains, cognition and workflow appear to interact recursively rather than independently. Expert cognition continuously shapes workflow execution through perception, interpretation, judgement, and decision-making. At the same time, workflow outcomes generate feedback that modifies future cognition.

This reciprocal relationship forms a continuous cycle of dynamic coupling. Expertise therefore emerges neither from cognition alone nor from workflow alone, but from their ongoing interaction.

This proposition challenges linear models of expertise development that assume knowledge acquisition directly produces expert performance. Instead, expertise appears to emerge through repeated cycles of cognitive--workflow interaction occurring throughout professional practice. Over time, these cycles generate increasing levels of sophistication, adaptability, and context sensitivity. Expertise should therefore be understood as a process of continuous emergence rather than a static achievement.

\paragraph{Proposition 3: Expert performance is an outcome rather than the essence of expertise.}
A recurring assumption within expertise research is that expertise can be identified through superior performance. While expert performance provides important evidence of expertise, the present analysis suggests that performance should not be equated with expertise itself.

Across the cases, observable performance represents the external manifestation of deeper cognitive processes. The master jade carver's craftsmanship, the film director's creative decisions, the educator's instructional effectiveness, and the leader's strategic judgement all emerge from underlying cognitive--workflow interactions. Performance therefore functions as an outcome of Workflow Cognition rather than its defining characteristic.

This distinction has important theoretical implications. If expertise is reduced to performance alone, then the mechanisms responsible for generating expertise remain obscured. Workflow Cognition shifts attention from observable outcomes towards the dynamic architecture responsible for producing those outcomes. Consequently, expertise should be investigated not only through what experts achieve, but also through how expert cognition continuously operates during practice.

\paragraph{Proposition 4: Human internal cognitive structures emerge through Workflow Cognition.}
The final proposition concerns the long-term consequences of repeated cycles of Workflow Cognition. The analysis suggests that expertise is not generated through isolated episodes of practice. Rather, expertise emerges through the longitudinal accumulation of cognitive--workflow interactions operating across months, years, and often decades of professional engagement.

Repeated cycles of perception, interpretation, judgement, decision-making, action, feedback, adaptation, and reflection gradually produce increasingly stable internal cognitive structures. These structures organise future cognition, guide decision-making, and enable experts to respond effectively to complex situations. Such structures may include:

\begin{itemize}
    \item professional schemas;
    \item domain-specific heuristics;
    \item mental models;
    \item aesthetic preferences;
    \item contextual judgement frameworks; and
    \item professional intuition.
\end{itemize}

Importantly, these structures remain adaptive rather than fixed. They continue to evolve through ongoing interaction with practice and experience.

This proposition provides an important bridge to the next stage of the AI+Expert framework. If expertise emerges through the long-term formation of internal cognitive structures, then understanding expertise requires investigation into how such structures develop, stabilise, and evolve across time. The next paper therefore extends the present framework by introducing \emph{Longitudinal Tacit Cognition}, which examines how repeated cycles of Workflow Cognition gradually crystallise into enduring yet evolving internal cognitive structures.

\paragraph{Summary of analytic propositions.}
Collectively, the preceding propositions support four central claims:

\begin{enumerate}
    \item Workflow Cognition can be identified across diverse domains of expertise.
    \item Expertise emerges through recursive coupling between cognition and workflow.
    \item Expert performance is an observable outcome rather than the essence of expertise.
    \item Human internal cognitive structures emerge through long-term cycles of Workflow Cognition.
\end{enumerate}

Together, these propositions provide theoretical support for a shift from possession-based views of expertise towards an architectural view, in which expertise is understood as an emergent cognitive phenomenon generated through dynamic cognitive--workflow interaction.

\section{Discussion}

The analytic propositions developed in this paper suggest that expertise may be more appropriately understood as a dynamic cognitive phenomenon than as a static accumulation of knowledge, skill, or experience. By introducing \emph{Workflow Cognition} as the underlying architecture of expertise, this paper offers a new perspective on several longstanding questions within expertise research, education, and human--AI systems.

The implications extend beyond the illustrative cases examined in this paper. If expertise emerges through the recursive coupling of cognition and workflow, then many prevailing assumptions regarding knowledge transmission, professional learning, and computational representation require reconsideration.

\subsection{Beyond Tacit Knowledge}

One of the most significant implications of Workflow Cognition concerns the role of tacit knowledge within expertise research. For more than half a century, tacit knowledge has served as one of the dominant explanations for expertise \cite{polanyi1966logic}. The concept successfully explains why expert judgement is often difficult to articulate and why apprenticeship remains effective in many professional domains.

However, tacit knowledge primarily describes a property of expertise rather than the mechanism that generates expertise. It explains that experts may know more than they can tell, but it does not fully explain how expert cognition becomes organised, enacted, and continuously transformed throughout practice.

Workflow Cognition offers a broader explanatory framework. Rather than focusing on what remains implicit, Workflow Cognition focuses on how expertise continuously operates through recursive interactions between cognition and workflow. Under this perspective, tacit knowledge becomes one observable manifestation of a deeper cognitive architecture. The difficulty of articulation is not the defining characteristic of expertise itself, but one consequence of the complex interactions through which expertise emerges.

Consequently, expertise should not be reduced to tacit knowledge. Instead, tacit knowledge should be understood as one component within a larger system of Workflow Cognition.

\subsection{Towards a Computational Theory of Expertise}

A second implication concerns the computational representation of expertise. Most contemporary approaches to modelling expertise focus on explicit knowledge, procedural rules, behavioural data, or performance outcomes. While such approaches may capture important aspects of expertise, they often struggle to represent the adaptive and evolving nature of expert cognition.

Workflow Cognition suggests that expertise may be more effectively represented as a dynamic architecture than as a static repository of knowledge. From this perspective, the computational challenge shifts from storing expert knowledge to modelling interactions between cognition and workflow. Future computational systems may therefore require representations capable of capturing:

\begin{itemize}
    \item perception patterns;
    \item interpretative frameworks;
    \item judgement structures;
    \item decision pathways;
    \item workflow adaptation processes; and
    \item longitudinal cognitive evolution.
\end{itemize}

This perspective provides a potential foundation for a computational theory of expertise in which expertise is represented as a living cognitive system rather than as a collection of informational assets.

However, modelling expertise as Workflow Cognition introduces a prior challenge. Before expert cognitive architectures can be computationally represented, they must first be systematically identified and translated. Within the broader AI+Expert framework, this challenge motivates the development of \emph{Questionnaire A: Cognitive Discovery Instrument}, a domain-agnostic instrument designed to identify expertise families, cognitive patterns, identity signals, and emerging expertise structures across diverse professional domains.

Rather than focusing solely on explicit knowledge, professional credentials, or observable performance, Questionnaire A seeks to reveal the underlying cognitive foundations through which expertise manifests. From this perspective, Questionnaire A functions as an initial discovery layer for computational expertise systems. It provides a structured mechanism for identifying potential Workflow Cognition patterns prior to computational modelling. More importantly, it establishes a systematic pathway through which Human Internal Cognitive Structures may become observable, analysable, and ultimately representable within computational environments.

The significance of such discovery instruments extends beyond expertise assessment. If expertise is fundamentally generated through Workflow Cognition, then computational systems require methods for capturing the cognitive--workflow structures that underlie expert judgement, decision-making, adaptation, and reflection. Without such discovery mechanisms, computational models remain limited to representing knowledge artefacts or behavioural outcomes rather than expertise itself.

\paragraph{Implications for AI Expert Twin systems.}
These implications are particularly significant for AI Expert Twin systems. If expertise can be systematically discovered through cognitive discovery instruments such as Questionnaire A, then AI representations of expertise no longer need to rely solely on static knowledge extraction. Instead, AI Expert Twins may seek to model the cognitive--workflow structures through which expertise operates.

This suggests a shift from knowledge-based AI towards cognition-based AI. Rather than simply reproducing expert answers, future AI Expert Twins may seek to reproduce the cognitive architecture underlying expert perception, interpretation, judgement, decision-making, adaptation, and reflection. Such systems would represent not only what experts know, but also how expertise continuously functions across practice.

From a Workflow Cognition perspective, the long-term objective of AI Expert Twins is therefore not merely the preservation of expert knowledge, but the computational representation of the cognitive architectures through which expertise emerges, evolves, and operates.

While the present paper focuses on the ontological foundations of expertise, future research will investigate how cognitive discovery instruments, Workflow Cognition representations, and AI Expert Twin systems may collectively contribute to a computational science of expertise.

\subsection{Implications for Education}

The educational implications of Workflow Cognition are equally significant. Traditional educational systems frequently prioritise content transmission. Learning is often conceptualised as the transfer of information from teacher to learner. Assessment systems similarly focus on the acquisition and reproduction of knowledge.

However, if expertise emerges through Workflow Cognition, then knowledge alone is insufficient for expertise development. Expertise requires learners to participate in recursive cycles of perception, interpretation, judgement, decision-making, action, feedback, and reflection. Educational processes must therefore support the formation of cognitive architectures rather than merely the acquisition of information.

This perspective shifts educational attention from content transmission towards cognitive transmission. The objective of education becomes not only helping learners know what experts know, but enabling learners to think, judge, decide, and adapt in ways that resemble expert cognition.

Such an approach aligns with apprenticeship traditions while also providing a theoretical foundation for future AI-assisted learning environments. From a Workflow Cognition perspective, the ultimate goal of education is not knowledge replication but cognitive reconstruction.

\subsection{Implications for Human--AI Systems}

The broader implications of Workflow Cognition extend to emerging human--AI systems. Many current AI systems focus primarily on information retrieval, pattern recognition, and content generation. While these capabilities are valuable, they remain limited in their ability to represent and reproduce expertise as a dynamic cognitive phenomenon.

Workflow Cognition suggests an alternative direction for human--AI collaboration. Rather than treating AI as a repository of information, future systems may function as cognitive infrastructures capable of supporting the representation, transmission, and reconstruction of expertise.

This perspective forms part of the broader AI+Expert framework. Within AI+Expert systems, human expertise is not viewed as static knowledge to be stored and retrieved. Instead, expertise is understood as an evolving cognitive architecture that can be translated, represented, reconstructed, and continuously refined through human--AI interaction.

Workflow Cognition therefore provides a theoretical bridge between expertise research and future computational systems designed to support expertise preservation, education, and large-scale transmission. More broadly, it suggests that the future of human--AI collaboration may depend less on replacing expertise and more on understanding, representing, and amplifying the cognitive architectures through which expertise emerges.

\section{Limitations and Future Research}

This paper is primarily theoretical in scope. Its objective is to develop Workflow Cognition as a conceptual framework for understanding expertise, rather than to provide a comprehensive empirical validation of the framework. The comparative expert cases are, therefore, used as illustrative cases that demonstrate the potential applicability of Workflow Cognition across diverse domains, rather than as a systematic empirical comparison.

A few limitations follow from this scope. First, the selected cases do not exhaust the range of professional domains in which expertise operates. Further work should examine whether similar cognitive--workflow structures can be observed across additional domains, including medicine, engineering, scientific research, entrepreneurship, cultural heritage practices, and other forms of professional judgement. Second, because Workflow Cognition proposes that expertise emerges through repeated cycles of dynamic coupling over extended periods of practice, longitudinal research is required to examine how these processes unfold across months, years, and decades of professional development. Third, while the framework proposes a dynamic cognitive architecture of expertise, further work is needed to determine how Thinking Flow, Workflow Evolution, Dynamic Coupling, and Expertise Emergence can be formally represented, measured, and computationally modelled.

These limitations define the next stage of the broader AI+Expert research programme. Workflow Cognition establishes an initial theoretical foundation for understanding how expertise operates. Future research will extend this foundation by examining how expertise forms, how it crystallises into distinctive cognitive structures, and how it may eventually be represented within computational systems.

\paragraph{Longitudinal Tacit Cognition.}
The first extension concerns the formation of expertise over time. While Workflow Cognition explains how cognition and workflow interact during professional practice, repeated cycles of such interaction may gradually produce increasingly stable internal cognitive structures that influence future perception, interpretation, judgement, and decision-making. Future work will therefore examine \emph{Longitudinal Tacit Cognition}, understood as the evolving internal cognitive structures formed through persistent cycles of Workflow Cognition across extended periods of professional engagement. The central question for this line of work is: how do Human Internal Cognitive Structures emerge through long-term cycles of Workflow Cognition?

This direction shifts attention from cognitive architecture to cognitive formation. It asks how professional schemas, domain-specific heuristics, mental models, intuitive judgement, and other internal structures are gradually formed, stabilised, and adapted through repeated engagement with practice.

\paragraph{Longitudinal Aesthetic Cognition.}
A second extension concerns the crystallisation of aesthetic and creative expertise. Not all internal cognitive structures develop in the same way. Some become associated with distinctive forms of aesthetic judgement, creative discernment, professional taste, and culturally recognised excellence. These forms of expertise often display stability and recognisability while remaining adaptive and generative.

Future research will therefore examine \emph{Longitudinal Aesthetic Cognition}, understood as the process through which certain internal cognitive structures become increasingly coherent, recognisable, and aesthetically distinctive through extended professional practice. The central question for this line of work is: how do internal cognitive structures crystallise into enduring forms of aesthetic expertise? This direction is particularly important for understanding expertise in domains such as craftsmanship, design, artistic production, cultural heritage, and other fields in which excellence is not reducible to technical correctness alone.

\paragraph{Expertise Workflow Grammar.}
A third extension concerns the computational representation of expertise. While Workflow Cognition explains how expertise operates, and Longitudinal Tacit Cognition explains how expertise forms, further work is required to examine how expertise may be translated, represented, and reconstructed within computational systems.

Future research will therefore develop \emph{Expertise Workflow Grammar}, a framework for identifying recurring cognitive--workflow patterns that may function as a representational language for expertise. The central question for this line of work is: can expertise be represented through a computable grammar derived from recurring cognitive and workflow structures? This direction provides a pathway towards computational expertise systems, AI Expert Twins, and future human--AI infrastructures capable of preserving, transmitting, and reconstructing expertise at scale.

Taken together, these future directions extend Workflow Cognition from a theory of expertise operation towards a broader computational science of expertise. The proposed progression is as follows:

\begin{itemize}
    \item \emph{Workflow Cognition} explains how expertise operates.
    \item \emph{Longitudinal Tacit Cognition} explains how expertise forms.
    \item \emph{Longitudinal Aesthetic Cognition} explains how expertise crystallises.
    \item \emph{Expertise Workflow Grammar} explains how expertise becomes computable.
\end{itemize}

This progression contributes to the long-term objective of the AI+Expert framework: transforming expertise from an implicit and often invisible human capability into a form of cognitive infrastructure that can be studied, represented, learned, and scaled.

\section{Conclusion}

This paper has argued that expertise should not be understood primarily as accumulated knowledge, acquired skill, professional experience, or tacit knowledge alone. While these perspectives explain important aspects of expert practice, they do not adequately account for the cognitive architecture through which expertise continuously emerges and evolves.

To address this limitation, the paper introduced \emph{Workflow Cognition} as a theoretical framework for understanding expertise. Workflow Cognition conceptualises expertise as a dynamic cognitive architecture generated through the recursive coupling of \emph{Thinking Flow} and \emph{Workflow Evolution}. From this perspective, expertise is not a static possession, but an ongoing cognitive phenomenon produced through continuous interaction between cognition and practice.

The paper further proposed \emph{Dynamic Coupling Theory} to explain how expertise develops through repeated cycles of perception, interpretation, judgement, decision-making, action, feedback, adaptation, and reflection. Rather than treating cognition and workflow as separate domains, Workflow Cognition positions them as mutually transformative processes operating throughout professional practice.

Building on this framework, the paper advanced a new ontological definition of expertise: expertise is an emergent manifestation of Workflow Cognition, continuously generated through the dynamic coupling of Thinking Flow and Workflow Evolution across time. Under this definition, knowledge, skills, decisions, aesthetic preferences, professional behaviours, and expert performance are not expertise itself. They are observable expressions of a deeper cognitive architecture through which expertise operates.

The comparative expert cases suggest that Workflow Cognition may be observable across diverse domains, including craftsmanship, creative production, education, and leadership. Despite substantial differences in context and professional activity, these cases point to recurring patterns of cognitive--workflow interaction. This supports the possibility that expertise may possess domain-specific manifestations while also resting upon domain-independent cognitive foundations.

Overall, the paper proposes a shift from possession-based theories of expertise towards an architectural view of expertise. Expertise is not simply something individuals possess; it is a continuously evolving cognitive system generated through the dynamic organisation of thinking and workflow. Workflow Cognition therefore provides a theoretical foundation for future investigations into expertise formation, aesthetic cognition, computational expertise, AI Expert Twins, and the broader AI+Expert framework.

\subsection*{Copyright Notice}

\noindent
\textcopyright\ 2026 Annie Yihong Yuan. All rights reserved. All figures, diagrams, interface examples, and visual materials in this paper are original works of the author unless otherwise stated.


\bibliographystyle{unsrtnat}
\bibliography{references}  






\end{document}